\newcommand{\be}{\begin{equation}}\newcommand{\ee}{\end{equation}}
\newcommand{\bea}{\begin{eqnarray}}\newcommand{\eea}{\end{eqnarray}}
\newcommand{\nn}{\nonumber}\newcommand{\p}[1]{(\ref{#1})}
\documentstyle[12pt]{article}

\begin{document}
\renewcommand{\thefootnote}{\fnsymbol{footnote}}
\thispagestyle{empty}
\hfill JINR E2-94-440 \\
\phantom{xxx} \hfill LNF-94/069(P)\\
\phantom{xxx} \hfill hep-th/9411168 \vspace{1.5cm} \\
\begin{center}
{\bf LINEARIZING $W_{2,4}$ AND $WB_2$  ALGEBRAS} \vspace{1.5cm} \\
S. Bellucci${}^a$\footnote{E-mail: bellucci@lnf.infn.it},
S. Krivonos${}^{a,b}$\footnote{E-mail: krivonos@thsun1.jinr.dubna.su}
and
A. Sorin${}^{a,b}$\footnote{E-mail: sorin@thsun1.jinr.dubna.su}
                    \vspace{1cm} \\
${}^a$INFN-Laboratori Nazionali di Frascati, P.O.Box 13 I-00044 Frascati,
            Italy \\
${}^b$Bogoliubov Theoretical Laboratory, JINR, Dubna, Russia
\vspace{2.5cm} \\
{\bf Abstract}
\end{center}
It has recently been shown that the $W_3$ and $W_3^{(2)}$ algebras
can be considered as subalgebras in some linear conformal algebras.
In this paper
we show that the nonlinear algebras $W_{2,4}$ and $WB_2$ as well as
Zamolodchikov's spin $5/2$ superalgebra also can be embedded as subalgebras
into some linear conformal algebras with a finite set of currents.
These linear algebras give rise to new realizations of the nonlinear
algebras which could be suitable in the construction of $W$-string theories.
\vspace{2.5cm} \\
\begin{center}
{\it Submitted to Phys. Lett. B}\\
\vfill
November 1994
\end{center}
\setcounter{page}0
\renewcommand{\thefootnote}{\arabic{footnote}}
\setcounter{footnote}0
\newpage

\section{Introduction}

In the last few years a lot of attention has been devoted to nonlinear
algebras. A great success was achieved in the construction of
$W$-algebras from the linear Kac-Moody (affine) algebras by
gauging some set of constraints imposed on the currents of the latter (see
Refs.[1] and references therein for reviews). In this way
Miura-like free-fields realizations [2] of $W$-algebras were also obtained.
This deep
connection between $W$- and affine algebras by no means implies that the
currents of former algebra can be constructed directly from the currents of
the latter one [3,4]. This remarkable relation between linear and
nonlinear algebras allows one to analyze the properties of
$W$-algebras and the corresponding $W$-strings from the
affine algebras point of view.

A different way to connect $W$-algebras to linear ones was proposed by
two of us [5]. We addressed the question, whether linear conformal Lie
algebras with
a finite number of currents exist, such that they contain nonlinear
W-algebras as subalgebras in some nonlinear basis?
It has been proved, for the simplest case of $W_3$ and $W_3^{(2)}$
algebras, that such linear algebras exist, have the structure of
an extension of the $W$-currents by some set of currents forming a
realization of the given $W$-algebra.

In an effort to show that the existence of the linear structure
is a general feature of the $W$-type algebras, rather than an
exception valid only for the simple cases mentioned above,
and motivated by the expectation that the properties of the
theories based on nonlinear algebras can be understood in the
simplest way from the corresponding linear structures,
we extend in the present paper the list of nonlinear algebras which
admit linearization. We analyze
from this point of view three $W$-type algebras:
$W_{2,4}$ algebra [6], the simplest fermionic nonlinear $WB_2$ algebra [7],
and Zamolodchikov spin $5/2$ algebra [8] which exists only for some particular
value of the central charge.
We find that in all cases the linear conformal algebras indeed exist and
are connected with the aforementioned types of $W$-realizations,
and contain the considered $W$-algebras as subalgebras. Using the
linear conformal algebras obtained we find, as a first application, a
new field realization of $W_{2,4}$ algebra.
At the level of the
linear algebras, the intrinsic relation between
the classical $W_{2,4}$ and $WB_2$ algebras is more transparent,
in the sense that the linear algebra for classical $W_{2,4}$
 is simply a subalgebra of the one for $WB_2$.

\section{$WB_2$ and $W_{2,4}$ algebras}

In this Section we fix our notation by displaying the explicit
structure of $WB_2$ [7] (or, using a different notation, $W(2,\frac{5}{2},4)$)
 and $W_{2,4}$ [6] algebras and their relations in both
the quantum and the classical cases.

The $WB_2$ algebra is the simplest example of a superalgebra with
a quadratic nonlinearity.
Besides the standard Virasoro stress-tensor $T(z)$
it contains two additional currents:
the bosonic current $U(z)$ with  spin $4$ and the fermionic current $Q(z)$
with spin $5/2$. This algebra exists for the generic central charge $c$ and
their currents obey the following
operator product expansions (OPE's)\footnote{As usual, we will write OPE's
in the quantum as well as in the classical case keeping in mind that for the
classical case only one pair contractions took place. The currents in the
r.h.s. of the OPE's are evaluated at the point $z_2$, and $z_{12}=z_1-z_2$.}:
\begin{eqnarray}
T(z_1)T(z_2) & = &  \frac{c/2}{z_{12}^4}+
        \frac{2T}{z_{12}^2}+\frac{T'}{z_{12}}\quad ,   \\
T(z_1)Q(z_2) & = &  \frac{5/2\;Q}{z_{12}^2}+\frac{Q'}{z_{12}} \quad ,\\
T(z_1)U(z_2) & = &  \frac{4U}{z_{12}^2}+\frac{U'}{z_{12}} \quad ,  \\
Q(z_1)Q(z_2) & = &  \frac{2c/5}{z_{12}^5}+
    \frac{2T}{z_{12}^3}+\frac{T'}{z_{12}^2}+
    \left[ b_1 T''+b_2 U +
               b_3 L_4 \right]\frac{1}{z_{12}} \quad , \\
U(z_1)Q(z_2) & = &  \frac{d_1 Q}{z_{12}^4}+
    \frac{d_2 Q'}{z_{12}^3}+
     \left[d_3 Q''+ d_4 (T\;Q) \right] \frac{1}{z_{12}^2}+ \nn \\
    & &  \left[d_5 Q'''+ d_6 (T'\;Q) +d_7 (T\;Q')\right] \frac{1}{z_{12}}
                        \quad , \nn \\
U(z_1)U(z_2) & = &  \frac{c/4}{z_{12}^8}+\frac{2T}{z_{12}^6}+
       \frac{T'}{z_{12}^5}+
       \left[ a_1 U +a_2 L_4 +a_3 T''\right] \frac{1}{z_{12}^4}+ \nn \\
     & &   \left[a_1/2 \; U' +a_2/2\; L_4' +a_4 T'''\right] \frac{1}{z_{12}^3}+
       \left[ a_5 U'' +a_6 L_4'' +a_7T'''' + \right. \nn \\
    & &  \left. a_8 L_{61}+a_9 L_{62}+a_{10}L_{63}+a_{11}L_{64}\right]
   \frac{1}{z_{12}^2}+
   \left[ a_{12} U''' +a_{13}L_4''' +a_{14}T''''' + \right. \nn \\
    & &  \left. a_{8}/2\;L_{61}'+a_{9}/2\;L_{62}'+
        a_{10}/2\;L_{63}'+a_{11}/2\; L_{64}'\right]
   \frac{1}{z_{12}} \label{WB2} ,
\end{eqnarray}
where the composites $L_4$ (spin 4), and $L_{61}-L_{64}$ (all
with spin 6) are defined by
\begin{eqnarray}
L_4 & =&  (T\;T) ,\nn \\
L_{61} & =&  (T\; L_4) , \quad
L_{62}  =  (T'\;T') , \quad
L_{63}  =  (T\; U) , \quad
L_{64}  =  (Q'\; Q)   \label{composites}
\end{eqnarray}
with normal ordering understood for the products of currents. The values of all
coefficients in the quantum and classical cases are given in Appendix.

The $W_{2,4}$ algebra contains only the bosonic currents
$\left\{ T(z),U(z) \right\}$
with spins $\left\{ 2,4 \right\}$. The current $U(z)$ is primary
with respect to the Virasoro stress-tensor $T$. Therefore the
only OPE we need, in order to specify this algebra, is $U(z_1)U(z_2)$.
This OPE has the same form
as \p{WB2} with $a_{11}=0$ and all other coefficients as given in Appendix.

We would like to stress that for both $WB_2$ and $W_{2,4}$ algebras we
use  the non standard definition of
composite currents \p{composites}, which are {\it non} primary with
respect to the corresponding stress-tensor $T(z)$, in contrast with the
papers [6,7] where primary composites have been used.
Despite the less evident structure of the coefficients in the OPE's
(2.4)-(2.5),
the use of non primary composites gives a significant speed up of
all calculations (at least twice as fast) which have been done with the
help of the Mathematica [9] package OPEdefs [10].

Let us summarize some peculiarities of these algebras.

First of all, let us
notice that in the quantum $WB_2$ algebra the coefficient
$b_2=\sqrt{\frac{6(13c+14)}{5c+22}}$ before
the spin 4 current $U(z)$ in the r.h.s (2.4)
vanishes, for the particular value
$c=-\frac{13}{14}$ of the central charge. This means that for this
value of the central charge
the currents $T(z)$ and $Q(z)$ form (modulo null-fields) the closed
quantum algebra which is the spin $\frac{5}{2}$
algebra of Zamolodchikov [8].

Secondly, we would like to stress that for the classical algebras
all coefficients in the OPE $U(z_1)U(z_2)$ for $WB_2$ and $W_{2,4}$
coincide (with the obvious exception of $a_{11}$ which vanishes
for $W_{2,4}$).
This means that the classical $W_{2,4}$ algebra can be obtained from
$WB_2$ by truncation, namely by dropping from OPE's the current
$Q(z)$ and all composites constructed from $Q(z)$.
In other words, in the
classical case the composite current $(Q'\;Q)$ transforms homogeneously
through itself and $Q(z)$, and so it can be consistently dropped,
together with $Q(z)$. In the quantum case the
situation changes drastically.  As shown in the Appendix, the
values of the coefficients $a_1-a_{14}$ for $WB_2$ and $W_{2,4}$ are
quite different and there exists no truncation procedure connecting these
algebras. The reason that singles out the quantum case is the following:
despite the existence of a primary spin 6 current which looks like
$(Q'\; Q) + \mbox{corrections} $, its quantum OPE with $U(z)$
{\it necessarily} contains
a term proportional to the $U(z)$ current in the right-hand side.
So, in the quantum $WB_2$ algebra we cannot exclude from the OPE's the
current $Q(z)$ and all composites constructed from $Q(z)$,
without contradicting the Jacobi identities.
That is why we need to consider the quantum $W_{2,4}$ algebra
independently from the $WB_2$ one.

In the next Section we construct the linear algebra which
contains the $W_{2,4}$ one as a subalgebra.

\setcounter{equation}0
\section{Linearization of $W_{2,4}$ algebra}

In this Section we show that there exists a linear algebra
$W_{2,4}^{lin}$ with a
finite number of current which contains $W_{2,4}$  as a subalgebra.

The main idea of our approach [5] is to extend the
set of currents of the given nonlinear algebra by adding some new currents in
such a way that
\begin{itemize}
\item the resulting set of currents still form a closed algebra,
\item the extended algebra contains the Virasoro subalgebra and all other
currents could be chosen to be  primary  with respect to Virasoro
stress-tensor,
\item there exists an {\em invertible} nonlinear transformation to
            another basis for the currents where all OPE's become linear.
\end{itemize}
In other words, here we would like to demonstrate that there is some
{\em linear, conformal} algebra which contains the $W_{2,4}$
algebra as a subalgebra in a certain nonlinear basis.

It is well known  [3,4], that the currents of the  nonlinear
algebra obtained through Drinfeld-Sokolov reduction, can be expressed in
terms of the corresponding  WZNW sigma-model currents. The novelties of our
approach are as follows:
\begin{itemize}
\item the number of currents in the linear algebra is the same as in the
    extended nonlinear one and the transformation from one set of currents to
    another  is invertible,
\item the linear algebra is conformal, i.e. it contains the Virasoro
  subalgebra and all other currents are primary  with respect to it.
\end{itemize}

Let us demonstrate that for the $W_{2,4}$ algebra the corresponding linear
algebra with the aforementioned properties does really exist.

The starting point of our construction is the following linear algebra
$W_{2,4}^{lin}$:
\begin{eqnarray}
T(z_1)T(z_2) & = &  \frac{c/2}{z_{12}^4}+
        \frac{2T}{z_{12}^2}+\frac{T'}{z_{12}}  \quad , \quad
T(z_1)J(z_2)  =   \frac{c_1}{z_{12}^3}+
                   \frac{J}{z_{12}^2}+\frac{J'}{z_{12}}\quad , \nn \\
T(z_1)U_1(z_2) & = &  \frac{4U_1}{z_{12}^2}+\frac{U_1'}{z_{12}} \quad , \quad
J(z_1)U_1(z_2)  =   \frac{2q_1 U_1}{z_{12}}\quad , \nn \\
J(z_1)J(z_2) & = &  \frac{1}{z_{12}^2} \quad , \quad
U_1(z_1)U_1(z_2)  =   \mbox{regular} \quad ,\label{W24l}
\end{eqnarray}
where the central charges $c$ and $c_1$ are connected with the
$U(1)$ charge $q_1$ as follows:
\begin{equation}
c=-\frac{120q_1^4-86q_1^2+15}{q_1^2} \quad , \quad
c_1=\frac{2-6q_1^2}{q_1} \; . \label{cc}
\end{equation}
Let us remark that we demand that the stress-tensor of the
nonlinear $W_{2,4}$
algebra coincides with the stress-tensor of the linear algebra \p{W24l}.
That is why we need to have
the central charge $c_1$ in the OPE $T(z_1)J(z_2)$.
Of course, we are free to pass by the transformation
$T\rightarrow T+\alpha J'$ to another basis
where all currents are primary. However,
in the basis under consideration all expressions look more tractable.

Now it is a matter of straightforward calculations to show that after the
following {\em invertible} nonlinear transformation to the new basis
$\{ J,T,U \}$, where
\begin{eqnarray}
U & = & U_1+z_1(T\;T)+z_2(T'\;J)+z_3(T\;J')+z_4(T\;J\;J)+z_5T''+ \nn \\
  &   & z_6 (J\;J\;J\;J)+z_7(J'\;J\;J)+z_8(J''\;J)+z_9(J'\;J')+z_{10}J'''
\; ,                  \label{UW24}
\end{eqnarray}
the currents $T$ and $U$ form the
$W_{2,4}$ algebra (2.1),(2.3),(2.5) with the central
charge $c$ parametrized by $q_1$ \p{cc}, provided the coefficients
$\{ z_1-z_{10}\}$ are chosen to be
\begin{eqnarray}
z_1 & = & \frac{(8-27q_1^2)(3-4q_1^2)}{84(1-8q_1^4)}a_{10} \nn \\
z_2 & = & \frac{(1-4q_1^2)(600q_1^4-452q_1^2+75)}{168q_1(1-8q_1^4)}a_{10} \nn
\\
z_3 & = & \frac{(1-2q_1^2)(600q_1^4-452q_1^2+75)}{84q_1(1-8q_1^4)}a_{10} \nn \\
z_4 & = & \frac{600q_1^4-452q_1^2+75}{84(8q_1^4-1)}a_{10} \nn \\
z_5 & = & \frac{1440q_1^8-1860q_1^6+865q_1^4-181q_1^2+15}
                             {168q_1^2(8q_1^4-1)}a_{10}\nn\\
z_6 & = & -\frac{z_4}{2} \nn \\
z_7 & = & \frac{(3q_1^2-1)(600q_1^4-452q_1^2+75)}
                             {42q_1(1-8q_1^4)}a_{10} \nn \\
z_8 & = & \frac{(32q_1^4-16q_1^2+3)(600q_1^4-452q_1^2+75)}
                             {336q_1^2(1-8q_1^4)}a_{10} \nn \\
z_9 & = & \frac{(48q_1^4-36q_1^2+7)(600q_1^4-452q_1^2+75)}
                             {672q_1^2(1-8q_1^4)}a_{10} \nn \\
z_{10} & = & \frac{(96q_1^6-68q_1^4+19q_1^2-3)(600q_1^4-452q_1^2+75)}
                                      {2016q_1^3(1-8q_1^4)}a_{10}
\end{eqnarray}
and $a_{10}$ is defined as in the Appendix:
$$
a_{10}=\frac{8(7c-115)}{2c+25}\sqrt{\frac{6}{(5c+22)(14c+13)}} \quad .
$$

To obtain the classical limit, we need to make the following renormalizations
of the $U(1)$ current $J$, central charge $c_1$, and $U(1)$ charge $q_1$:
\begin{equation}
J \rightarrow \frac{1}{\sqrt{c}}J \quad , \quad
c_1 \rightarrow \sqrt{c}c_1 \quad , \quad
q_1 \rightarrow \sqrt{c}q_1 \quad , \quad
\end{equation}
and consider, as usual, the values of all coefficients in the limit
\begin{equation}
c\rightarrow \infty \quad \mbox{or} \quad q_1\rightarrow 0 \quad .
\end{equation}
The classical linear algebra ${\cal W}_{2,4}^{lin}$  has the
following form\footnote{We will use the calligraphic and tilded letters, to
distinguish the classical currents and charges from the corresponding
quantum ones.}:
\begin{eqnarray}
{\cal T}(z_1){\cal T}(z_2) & = &  \frac{{\tilde c}/2}{z_{12}^4}+
        \frac{2{\cal T}}{z_{12}^2}+\frac{{\cal T}'}{z_{12}}  \quad , \quad
{\cal T}(z_1){\cal J}(z_2)  =   \frac{-2i{\tilde c}/\sqrt{15}}{z_{12}^3}+
                   \frac{{\cal J}}{z_{12}^2}+\frac{{\cal J}'}{z_{12}}
           \quad ,  \nn \\
{\cal T}(z_1){\cal U}_1(z_2) & = &  \frac{4{\cal U}_1}{z_{12}^2}+
                         \frac{{\cal U}_1'}{z_{12}} \quad , \quad
{\cal J}(z_1){\cal U}_1(z_2)  = \frac{2i\sqrt{15}{\cal U}_1}{z_{12}}
                          \quad , \nn\\
{\cal J}(z_1){\cal J}(z_2) & = &  \frac{\tilde c}{z_{12}^2}\quad , \quad
{\cal U}_1(z_1){\cal U}_1(z_2)  = \mbox{regular} \quad .\label{W24lcl}
\end{eqnarray}
The classical expression for the current $\cal U$, forming together
with $\cal T$ the classical ${\cal W}_{2,4}$ algebra, reads
\begin{eqnarray}
{\cal U} & = & {\cal U}_1+\frac{24}{{\tilde c}\sqrt{105}}{\cal T}^2-
\frac{5i}{2{\tilde c}\sqrt{7}}{\cal T}'{\cal J}-
\frac{5i}{{\tilde c}\sqrt{7}}{\cal T}{\cal J}'-
\frac{75}{{\tilde c}^2\sqrt{105}}{\cal T}{\cal J}^2+
\frac{1}{2\sqrt{105}}{\cal T}''+ \nn \\
  &   & \frac{75}{2{\tilde c}^3\sqrt{105}}{\cal J}^4+
\frac{10i}{{\tilde c}^2\sqrt{7}}{\cal J}'{\cal J}^2-
\frac{15}{4{\tilde c}\sqrt{105}}{\cal J}''{\cal J}-
\frac{5}{8{\tilde c}}\sqrt{\frac{7}{15}}{\cal J}'{\cal J}'-
\frac{i}{24\sqrt{7}}{\cal J}'''
\; .                  \label{UW24cl}
\end{eqnarray}

So, we have explicitly shown that $W_{2,4}^{lin}$ algebra \p{W24l}
(${\cal W}_{2,4}^{lin}$  \p{W24lcl}) contains the quantum (classical)
$W_{2,4}$ algebra as its subalgebra in the nonlinear basis.

We postpone the discussion of the consequences of the linearization property
to Section 5. In the next Section we show how to construct the
linear algebra for the $WB_2$ superalgebra.

\setcounter{equation}0
\section{Linearization of $WB_2$ algebra}

In this Section we explicitly demonstrate that $WB_2$ is a
subalgebra of a very simple {\em linear} conformal superalgebra.

The main novelties of $WB_2$ algebra are, in comparison with $W_{2,4}$,
\begin{itemize}
\item the fermionic nature of the spin $5/2$ current $Q(z)$
\item the appearance of the spin $4$ current $U(z)$ in the OPE of two
        spin $5/2$ fermionic currents.
\end{itemize}
The last property means that the whole structure of $WB_2$ algebra is
encoded in the structure of the ``supersymmetry'' current $Q(z)$.
Moreover, the linear algebra $WB_2^{lin}$ we are going to construct
must reproduce at $c=-\frac{13}{14}$ the linear algebra for the Zamolodchikov's
spin $5/2$ nonlinear algebra.
In addition, we know that the classical ${\cal W}_{2,4}$ algebra is a
truncation of the classical ${\cal WB}_2$. In the case of the corresponding
linear algebras, due to the absence of nonlinear composites,
the ${\cal W}_{2,4}^{lin}$ \p{W24lcl} must be a bosonic subalgebra of
${\cal WB}_2^{lin}$.

Without going into details, let us show that $WB_2^{lin}$
contains the bosonic $\{ T,J,U_1 \}$ and fermionic $\{ S, Q_1 \}$
currents\footnote{We hope that the use of the same
notation for the bosonic currents as in the previous Section does not give
rise to confusion as to which algebra we are dealing with.},
obeying  the following OPE's\footnote{For the same reason of simplicity
as in the previous Section we work in the non-primary basis.}:
\begin{eqnarray}
T(z_1)T(z_2) & = &  \frac{c/2}{z_{12}^4}+\frac{2T}{z_{12}^2}+
                  \frac{T'}{z_{12}}  \quad , \nn \\
T(z_1)J(z_2) & = & \frac{c_2}{z_{12}^2}+ \frac{J}{z_{12}^2}+
                  \frac{J'}{z_{12}} \quad , \quad
T(z_1)S(z_2)  =   \frac{1/2 S}{z_{12}^2}+\frac{S'}{z_{12}} \quad  ,\nn \\
T(z_1)Q_1(z_2) & = & \frac{5/2 Q_1}{z_{12}^2}+\frac{Q_1'}{z_{12}} \quad , \quad
T(z_1)U_1(z_2)  =   \frac{4 U_1}{z_{12}^2}+\frac{U_1'}{z_{12}} \quad , \nn \\
J(z_1)Q_1(z_2) & = &  \frac{q_2 Q_1}{z_{12}} \quad , \quad
J(z_1)U_1(z_2)  =   \frac{2q_2 U_1}{z_{12}} \quad ,\nn \\
S(z_1)S(z_2) & = &  \frac{1}{z_{12}^2}  \quad , \quad
J(z_1)J(z_2)  =   \frac{1}{z_{12}^2} \quad , \quad
Q_1(z_1)Q_1(z_2)  =  \frac{b_2 U_1}{z_{12}}  \label{WB2lin},
\end{eqnarray}
where the central charges $c$ and $c_2$ are parametrized by the $U(1)$ charge
$q_2$
\begin{equation}
c=-\frac{120q_2^4-125q_2^2+30}{2q_2^2} \quad , \quad
c_2=\frac{2-4q_2^2}{q_2} \; , \label{ccwb}
\end{equation}
and the coefficient $b_2$ defined in the Appendix reads
$$
b_2=\sqrt{\frac{6(14c+13)}{5c+22}} \quad .
$$

In order to prove this, we do the following {\em invertible} transformation
to the new basis $\{T,J,S,Q,U \}$, where
\begin{equation}
Q  = Q_1+z_1 S''+z_2 (J\; S') + z_3 (J'\; S)+
     z_4 (J\; J\;S)+z_5 (T \; S) \label{newbas}
\end{equation}
and $U$ is defined from the OPE in eqs. (2.4)
$$
Q(z_1)Q(z_2) =   \frac{2c/5}{z_{12}^5}+
    \frac{2T}{z_{12}^3}+\frac{T'}{z_{12}^2}+
    \left[ b_1 T''+b_2 U + b_3 (T\;T) \right]\frac{1}{z_{12}} \; .
$$
If the parameters $\{z_1-z_5\}$ are chosen to be
\begin{eqnarray}
z_1 & = & \frac{8q_2^4-5q_2^2+2}{4q_2^2}\sqrt{\frac{30}{25+2c}} \nn \\
z_2 & = & \frac{2q_2^2-1}{q_2}\sqrt{\frac{30}{25+2c}} \nn \\
z_3 & = & \frac{6q_2^2-3}{2q_2}\sqrt{\frac{30}{25+2c}} \nn \\
z_4 & = & \sqrt{\frac{30}{25+2c}} \nn \\
z_5 & = & -\sqrt{\frac{30}{25+2c}}    \label{coef}
\end{eqnarray}
and the central charge $c$ is connected with $q$ as in \p{ccwb}, then the
currents $T,Q$ and $U$ just form the $WB_2$ algebra \p{WB2}.

Thus, one concludes that $WB_2^{lin}$ indeed contains $WB_2$ as a subalgebra.

Let us close this Section with some comments.

First of all, we would like to remark that the linear $WB_2^{lin}$
algebra \p{WB2lin} and the coefficients $\{z_1-z_5\}$ \p{coef} do not
contain singularities when $c\rightarrow -13/14$ ($q\rightarrow \sqrt{5/14}$).
Therefore, we can immediately read off the linear algebra for the
Zamolodchikov's spin $5/2$ algebra with  $c=-13/14$
\begin{eqnarray}
T(z_1)T(z_2) & = &  \frac{-13/28}{z_{12}^4}+\frac{2T}{z_{12}^2}+
                  \frac{T'}{z_{12}} \quad , \nn \\
T(z_1)J(z_2) & = & \frac{4\sqrt{2/35}}{z_{12}^2}+ \frac{J}{z_{12}^2}+
                  \frac{J'}{z_{12}} \quad , \quad
T(z_1)S(z_2)  =   \frac{1/2 S}{z_{12}^2}+\frac{S'}{z_{12}}\quad , \nn \\
T(z_1)Q_1(z_2) & = & \frac{5/2 Q_1}{z_{12}^2}+\frac{Q_1'}{z_{12}}\quad , \quad
J(z_1)Q_1(z_2)  =   \frac{\sqrt{5/14} Q_1}{z_{12}}\quad , \nn \\
S(z_1)S(z_2) & = &  \frac{1}{z_{12}^2}  \quad , \quad
J(z_1)J(z_2)  =   \frac{1}{z_{12}^2} \quad , \quad
Q_1(z_1)Q_1(z_2)  =  \mbox{regular}  \label{W5/2}.
\end{eqnarray}
The corresponding expression for the spin $5/2$ current $Q_Z$, which together
with the stress-tensor $T$ forms the Zamolodchikov's spin $5/2$ algebra,
reads as follows:
\begin{equation}
Q_Z  = Q_1+\frac{121}{12\sqrt{105}} S''-\frac{2}{3}\sqrt{\frac{2}{3}} (J\; S')
- \sqrt{\frac{2}{3}} (J'\; S)+
     \frac{1}{3}\sqrt{\frac{35}{3}} (J\; J\;S)-
\frac{1}{3}\sqrt{\frac{35}{3}} (T \; S) \label{zam}
\end{equation}
Thus we showed that the linear superalgebra \p{W5/2} contains the
Zamolodchikov's spin $5/2$ as a subalgebra.

Secondly, as pointed out in the Section 2, the classical
${\cal WB}_2$ algebra contains the classical ${\cal W}_{2,4}$ algebra, in
the truncation limit $Q\rightarrow 0$.
It can be easily shown that the analogous relation exists
also for the linear algebra ${\cal WB}_2^{lin}$. In the classical
limit $c\rightarrow \infty$, after the redefinitions
\begin{equation}
J \rightarrow \frac{1}{\sqrt{c}}J \quad , \quad
S \rightarrow \frac{1}{\sqrt{c}}S \quad , \quad
c_2 \rightarrow \sqrt{c}c_2 \quad , \quad
q_2 \rightarrow \sqrt{c}q_2 \quad ,
\end{equation}
the classical linear algebra ${\cal WB}_2^{lin}$ has the
following form:
\begin{eqnarray}
{\cal T}(z_1){\cal T}(z_2) & = &  \frac{{\tilde c}/2}{z_{12}^4}+
       \frac{2{\cal T}}{z_{12}^2}+
                  \frac{{\cal T}'}{z_{12}}\quad , \nn \\
{\cal T}(z_1){\cal J}(z_2) & =&  \frac{-2i{\tilde c}/\sqrt{15}}{z_{12}^2}+
       \frac{{\cal J}}{z_{12}^2}+
                  \frac{{\cal J}'}{z_{12}} \quad , \quad
{\cal T}(z_1){\cal S}(z_2)  =   \frac{1/2 {\cal S}}{z_{12}^2}+
                   \frac{{\cal S}'}{z_{12}}\quad , \nn \\
{\cal T}(z_1){\cal Q}_1(z_2) & = & \frac{5/2 {\cal Q}_1}{z_{12}^2}+
                   \frac{{\cal Q}_1'}{z_{12}} \quad , \quad
{\cal T}(z_1){\cal U}_1(z_2)  = \frac{4 {\cal U}_1}{z_{12}^2}+
                   \frac{{\cal U}_1'}{z_{12}}\quad , \nn \\
{\cal J}(z_1){\cal Q}_1(z_2) & = &  \frac{i\sqrt{15} {\cal Q}_1}{z_{12}}
        \quad , \quad
{\cal J}(z_1){\cal U}_1(z_2)=\frac{2i\sqrt{15}{\cal U}_1}{z_{12}}\quad ,\nn \\
{\cal S}(z_1){\cal S}(z_2) & = &  \frac{\tilde c}{z_{12}^2}\quad ,\quad
{\cal J}(z_1){\cal J}(z_2)  =  \frac{\tilde c}{z_{12}^2} \quad , \quad
{\cal Q}_1(z_1){\cal Q}_1(z_2) = \frac{2\sqrt{21/5} {\cal U}_1}{z_{12}} .
                    \label{WB2lincl}
\end{eqnarray}
One can immediately see that the bosonic subalgebra of ${\cal WB}_2^{lin}$
coincides with  ${\cal W}_{2,4}^{lin}$ \p{W24lcl}. While the
classical $W_{2,4}$ algebra is a truncation of $WB_2$, the corresponding
linear algebra $W_{2,4}^{lin}$ is a simple subalgebra of $WB_2^{lin}$.

In the next Section we consider some examples that can illustrate
the use of the linearized
algebra for both $W_{2,4}$ and $WB_2$ algebras.

\setcounter{equation}0
\section{Conclusions and discussion}
In this paper we extend the list of nonlinear algebras which admit
linearization, by explicitly constructing the linear conformal algebras,
with a finite set of currents, which contain $W_{2,4}$ and $WB_2$
algebras as subalgebras in a nonlinear basis. Thus, now we know that
the first representatives of both the $WA_n$ $(W_3$ [5]) and the
$WB_n$ $(WB_2)$ series of nonlinear algebras can be linearized.
Despite the lack, up to now, of a general algorithmic procedure to
construct the linear conformal algebra for any given nonlinear algebra,
it makes sense to conjecture that the possibility of linearization is a
general property
of nonlinear algebras rather than a peculiarity of some exceptional cases.

Let us finish by discussing some common properties of the
linearization and the linear algebras.

First of all, we would like to note that, in order to linearize a nonlinear
algebra,
we need to extend it by adding some currents ($J(z)$ in the case of
$W_{2,4}$ and $\{ J(z),S(z) \}$ in the case of $WB_2$ algebra). Thus,
the first open question is as to which additional properties of the system
are associated with these symmetries.

Secondly, it is interesting that the constructed linear algebras are
{\it homogeneous} with respect to some currents ($U_1(z)$ and
$\{ U_1(z),Q_1(z) \}$ in the case of $W_{2,4}$ and $WB_2$, respectively).
This property means that we could consistently put these currents
equal to zero and be left with the realization of the algebras in terms of
an arbitrary stress-tensor and a spin 1 current $J(z)$ for $W_{2,4}$,
and a spin 1 current $J(z)$ and a spin $1/2$ current $S(z)$ for the
$WB_2$.

Finally, let us remark that, owing to the invertible relation between
the currents of the nonlinear and the linear algebras, every realization
of $W_{2,4}^{lin}$ or $WB_2^{lin}$ is a realization of $W_{2,4}$
or $WB_2$ respectively. So, the problem of constructing the
realizations of these algebras is reduced to the problem of constructing
realizations of the linear algebras $W_{2,4}^{lin}$ and $WB_2^{lin}$.
In the rest of this Section we will present an example of such realization
for the case of $W_{2,4}$ algebra.

{}From the simple structure of the $W_{2,4}^{lin}$ algebra (3.1)
it is clear that its most general realization includes at least
two scalar fields $\phi_{i}$ $(i=1,2)$ with OPE's
\begin{equation}
\phi_i (z_1)\phi_j (z_2) = -\delta_{ij} \ln (z_{12})
\end{equation}
and, commuting with them, a Virasoro stress tensor $\tilde T$ having a nonzero
central charge which we will denote as $c_T$. Representing the bosonic
primary current $U_1(z)$ in the standard way by an exponential of $\phi_i$,
we find the following expressions:
\begin{eqnarray}
T & = & {\tilde T} -\frac{1}{2}(\phi_1')^2-\frac{1}{2}(\phi_2')^2-
       \frac{i(1-3q_1^2)}{q_1}\phi_1''- \nn \\
  &   & \frac{i(4+12q_1^2-N)}{2\sqrt{N-4q_1^2}}\phi_2'' \quad , \nn \\
J & = & i\phi_1' \quad , \nn \\
U_1&= & s\; \mbox{exp}\left( i\sqrt{N-4q_1^2}\phi_2+2iq_1\phi_1\right)
          \quad , \nn \\
c_T & = & 3\left( \frac{(4-N+12q_1^2)^2}{N-4q_1^2}-
              \frac{(2q_1^2-1)^2}{q_1^2} \right) \quad , \label{realiz}
\end{eqnarray}
where $N$ runs over non-negative natural numbers and $s$ is an arbitrary
parameter (due to a $U_1$ rescaling invariance of the OPE's (3.1) $s$
can be chosen  to be 0 or 1).

In the case of $s=0$ the field $\phi_2$ can be absorbed by the stress-tensor
and we end with the following $\phi_2$ independent realization:
\begin{eqnarray}
T & = & {\tilde T}_0 -\frac{1}{2}(\phi_1')^2-
       \frac{i(1-3q_1^2)}{q_1}\phi_1'' \quad ,\nn \\
J & = & i\phi_1' \quad , \nn \\
U_1&= & 0\quad , \nn \\
c_{T_0} & = & 1-\frac{3(2q_1^2-1)^2}{q_1^2} \quad . \label{realiz1}
\end{eqnarray}

After substituting eqs. \p{realiz1} into (3.3), we get the known realization
of $W_{2,4}$ algebra [11], whereas the use of eqs. \p{realiz} in (3.3)
yields a new realization of $W_{2,4}$ algebra, which may play a role
for $W_{2,4}$ string theory [11].

A last comment concerns the realization \p{realiz1}, for which
the $W_{2,4}^{lin}$ algebra is  reduced to a direct product of a $U(1)$
algebra and the Virasoro one with the central charge $c_{T_0}$ \p{realiz1}.
Then, the minimal Virasoro models [12] which correspond to the
central charge
\begin{equation}
c_m=1-6\frac{(p-q)^2}{pq} \Rightarrow q_1^2 =\frac{p}{2q}
\end{equation}
give rise to the following induced central charges (3.2):
\begin{equation}
c_{W_{2,4}}^{min}= -2\frac{(5(2p)-6q)(3(2p)-5q)}{(2p)q}
\end{equation}
They coincide with the central charges of $W_{2,4}$ minimal models
[13]
\begin{equation}
c^{min}= -2\frac{(5{\tilde p}-6{\tilde q})
      (3{\tilde p}-5{\tilde q})}{{\tilde p}{\tilde q}}
\end{equation}
for even values of $\tilde p$.
\vspace{1cm}\\
\noindent{\large\bf Acknowledgments}

We are grateful  to L. Bonora, A. Honecker, K. Hornfeck,  E. Ivanov,
W. Nahm, V. Ogievetsky and S. Sciuto for many useful discussions.

Two of us (S.K. and A.S.) wish to thank INFN for financial support and
the Laboratori Nazionali di
Frascati for hospitality during the course of this work.
\newpage

\noindent{\large\bf Appendix}

Here we write down the expressions for the coefficients in the OPE's for
$WB_2$ and $W_{2,4}$ algebras.\vspace{0.5cm}\\

\begin{tabular}{|l|l|l||l|l|} \hline \hline
Coeff. & Quantum $WB_2$ & Classical $WB_2$ & Quantum $W_{2,4}$
        & Classical $W_{2,4}$ \\ \hline
$a_1$  & $\frac{3(2c^2+83c-490)}{R_1}\sqrt{\frac{6}{R_2R_3}}$
       & $3\sqrt{\frac{3}{35}}$
       & $3\sqrt{6}R_4 $
       & $3\sqrt{\frac{3}{35}}$ \\
$a_2$  & $\frac{42}{R_2}$
       &$ \frac{42}{5c}$
       & $ \frac{42}{R_2}  $
       &$ \frac{42}{5c}$ \\
$a_3$  & $\frac{3(c-4)}{2R_2}$
       & $ \frac{3}{10}$
       & $\frac{3(c-4)}{2R_2} $
       &$ \frac{3}{10}$ \\
$a_4$  & $\frac{2c-29}{6R_2}$
       &$ \frac{1}{15}$
       & $\frac{2c-29}{6R_2} $
       &$ \frac{1}{15}$ \\
$a_5$  & $\frac{10c^2-197c-2810}{2R_1}\sqrt{\frac{1}{6R_2R_3}}$
       &$ \frac{1}{4}\sqrt{\frac{5}{21}}$
       &$ \frac{5c+64}{2(c+24)\sqrt{6}}R_4 $
       &$ \frac{1}{4}\sqrt{\frac{5}{21}}$ \\
$a_6$  & $\frac{352c^2-704c-9125}{2R_1R_2R_3}$
       & $\frac{44}{35c}$
       & $\frac{176c^2+117c-2528}{2(2c-1)(7c+68)R_2} $
       & $\frac{44}{35c}$ \\
$a_7$  & $\frac{20c^3-462c^2-2862c+9865}{12R_1R_2R_3}$
       & $\frac{1}{84}$
       &$ \frac{10c^3-166c^2-1645c+2296}{12(2c-1)(7c+68)R_2} $
       & $\frac{1}{84}$ \\
$a_8$  & $\frac{108(32c-5)}{R_1R_2R_3}$
       & $\frac{864}{35c^2}$
       &$ \frac{24(72c+13)}{(2c-1)(7c+68)R_2}$
       & $\frac{864}{35c^2}$ \\
$a_9$  & $\frac{-3(38c^2-1669c-4930)}{2R_1R_2R_3}$
       & $\frac{-57}{140c}$
       & $ \frac{-3(19c^2-786c-2368)}{2(2c-1)(7c+68)R_2} $
       & $\frac{-57}{140c}$ \\
$a_{10}$  & $\frac{8(7c-115)}{R_1}\sqrt{\frac{6}{R_2R_3}}$
          &$ \frac{4}{c}\sqrt{\frac{21}{5}}$
          & $\frac{28\sqrt{6}}{c+24}R_4 $
          &$ \frac{4}{c}\sqrt{\frac{21}{5}}$ \\
$a_{11}$  & $\frac{60R_2}{R_1R_3}$
          &$ \frac{75}{7c}$
          & $0 $
          & $0 $ \\
$a_{12}$  & $\frac{2c^2-223c-670}{2R_1}\sqrt{\frac{1}{6R_2R_3}}$
          &$ \frac{1}{4\sqrt{105}}$
          & $\frac{c-4}{2(c+24)\sqrt{6}}R_4 $
          &$\frac{1}{4\sqrt{105}} $ \\
$a_{13}$  & $\frac{3(13c^2-278c-950)}{R_1R_2R_3}$
          &$ \frac{39}{140c}$
          & $\frac{3(13c^2-131c-342)}{2(2c-1)(7c+68)R_2} $
          &$ \frac{39}{140c}$ \\
$a_{14}$  & $\frac{20c^3-1196c^2+2953c+44150}{80R_1R_2R_3}$
          &$ \frac{1}{560}$
          & $\frac{10c^3-283c^2-898c+5296}{80(2c-1)(7c+68)R_2} $
          &$ \frac{1}{560}$ \\ \hline \hline
$b_1$     &  $ \frac{3(c-1)}{2R_2} $
          & $ \frac{3}{10}$
          &  --- & --- \\
$b_2$     & $\sqrt{\frac{6R_3}{R_2}} $
          &  $ 2\sqrt{\frac{21}{5}} $
          & --- &  --- \\
$b_3$     & $\frac{27}{R_2} $
          &  $ \frac{27}{5c} $
          & --- &  --- \\ \hline \hline
$d_1$     & $\frac{5}{4}\sqrt{\frac{3R_3}{2R_2}}$
          & $\frac{1}{4}\sqrt{105} $
          & ---  & --- \\
$d_2$     & $ \sqrt{\frac{3R_3}{2R_2}}$
          & $\sqrt{\frac{21}{5}} $
          & ---  & --- \\
$d_3$     & $\frac{5(c+8)}{2R_1}\sqrt{\frac{R_3}{6R_2}}$
          & $\frac{1}{4}\sqrt{\frac{35}{3}} $ & ---
            & --- \\
$d_4$     & $\frac{15}{R_1}\sqrt{\frac{3R_3}{2R_2}}$
          & $\frac{3}{2c}\sqrt{105}$  & ---
            & --- \\
$d_5$     & $\frac{2c-5}{R_1}\sqrt{\frac{R_2}{6R_3}}$
          & $\frac{1}{2}\sqrt{\frac{5}{21}}$  & ---
            & --- \\
$d_6$     & $\frac{5(22c-49)}{R_1}\sqrt{\frac{3}{2R_2R_3}}$
          & $\frac{11}{2c}\sqrt{\frac{15}{7}}$  & ---
            & --- \\
$d_7$     & $\frac{82c+215}{R_1}\sqrt{\frac{6}{R_2R_3}}$
          & $\frac{41}{c}\sqrt{\frac{3}{35}}$  & ---
            & --- \\ \hline
\end{tabular}
\phantom{xxx} \vspace{0.5cm}\\
Here we set

$$
R_1=2c+25,R_2=5c+22,R_3=14c+13,R_4=
        \sqrt{ \frac{(c+24)(c^2-172c+196)}{(5c+22)(7c+68)(2c-1)}}.
$$


\newpage
\begin{thebibliography}{99}
\bibitem{a1} P. Bouwknegt and K. Schoutens, Phys. Rep. 223 (1993) 183;\\
L. Feher, L. O'Raifeartaigh, P. Ruelle, I. Tsutsui and A. Wipf,
     Phys. Rep. 222 (1992) 1.
\bibitem{a2} A.B. Zamolodchikov and V.A. Fateev, Nucl. Phys. B280 (1987) 644;\\
             V.A. Fateev and S.L. Lukyanov, Int. J. Mod. Phys. A3 (1988) 507,\\
             Sov. Scient. Rev. A15 (1990), Sov. J. Nucl. Phys. 49 (1989) 925;\\
             A. Bilal and J.-L. Gervais, Nucl. Phys. B314 (1989) 646,
                                         Nucl. Phys B318 (1989) 579.
\bibitem{a3} F.A. Bais, T. Tjin and P. van Driel, Nucl. Phys. B357 (1991)
632;\\
O'Raifeartaigh, P. Ruelle, I. Tsutsui and A. Wipf,
     Commun. Math. Phys.  143 (1992) 333.
\bibitem{DeBoer} B.L. Feigin and E. Frenkel, Phys. Lett. B246 (1990) 75; \\
A. Sevrin and W. Troost, Phys. Lett. B315 (1993) 304;\\
J. de Boer and T. Tjin, Commun. Math. Phys.  160 (1994) 317.
\bibitem{sk1} S. Krivonos and A. Sorin, Phys. Lett. B335 (1994) 45.
\bibitem{a4} K.-J. Hamada and M. Takao, Phys. Lett. B209 (1988) 247;\\
             P. Bouwknegt, Phys. Lett. B207 (1988) 295;\\
             D.-H. Zhang, Phys. Lett. B232 (1989) 323;\\
             R. Blumenhagen, M. Flohr, A. Kliem, W. Nahm, A. Recknagel and
             R. Varnhagen, Nucl. Phys. B361 (1991) 255;\\
             H.G. Kausch and G.M.T. Watts, Nucl. Phys. B354 (1991) 740.
\bibitem{a5} J.M. Figueroa-O'Farrill, S. Schrans and K. Thielemans,
             Phys. Lett. B263 (1991) 378.
\bibitem{Z52} A. Zamolodchikov, Theor. Math. Phys. 65 (1985) 347.
\bibitem{Math} S. Wolfram,Mathematica, (Addison-Wesley, Reading, MA,1991).
\bibitem{Thielemans} K.Thielemans, Int. J. Mod. Phys. C2 (1991) 787.
\bibitem{Pope} H. Lu, C.N. Pope, X.J. Wang and S.C. Zhao, Class. Quantum
           Grav. 11 (1994) 939.
\bibitem{BPZ} A. Belavin, A. Polyakov and A. Zamolodchikov,
              Nucl. Phys. B280 (1984) 333.
\bibitem{a6} H.G. Kausch and G.M.T. Watts, Int. J. Mod. Phys. A7 (1992) 4175;\\
             W. Eholzer, M. Flohr, A. Honecker, R. Hubel, W. Nahm and
             R. Varnhagen, Nucl. Phys. B383 (1992) 249.
\end{thebibliography}
\end{document}